\documentclass{aa}
\usepackage{txfonts}
\usepackage{graphicx}
\usepackage{natbib}
\usepackage{amstext,cases}
\bibpunct{(}{)}{;}{a}{}{,}

\usepackage[x11names]{xcolor}

\begin{document}

\title{Crucial aspects of the initial mass function (II)}
\subtitle{The inference of total quantities from partial information on a cluster}

\author{Miguel Cervi{\~n}o$^{1,2}$,  Carlos Rom{\'a}n-Z{\'u}{\~n}iga$^{3}$, Amelia Bayo$^{4,5}$, Valentina Luridiana$^{2,6}$, N{\'e}stor S{\'a}nchez$^7$ and Enrique P{\'e}rez$^1$}
\institute{Instituto de Astrof{\'\i}sica de Andaluc{\'\i}a (IAA-CSIC), Glorieta de la Astronom{\'\i}a s/n, 18008 Granada, Spain
\and Instituto de Astrof{\'{\i}}sica de Canarias, c/ v{\'{\i}}a L\'actea s/n, 38205 La Laguna, Tenerife, Spain 
\and Instituto de Astronom{\'{\i}}a, Universidad Acad\'emica en Ensenada, Universidad Nacional Aut\'onoma de M\'exico,
Ensenada BC, 22860 Mexico
\and European Southern Observatory, Casilla 19001, Santiago 19, Chile
\and Max Planck Institut f\"ur Astronomie, K\"onigstuhl 17, 69117, Heidelberg, Germany
\and Departamento de Astrof{\'{\i}}sica, Universidad de La Laguna (ULL), 38205 La Laguna, Tenerife, Spain
\and S. D. Astronom\'{\i}a y Geodesia,
Fac. CC. Matem\'aticas,
Universidad Complutense de Madrid,
28040, Madrid,
Spain.
}
\offprints{M. Cervi\~no \email{mcs@iaa.es}}
\date{Received ; accepted }

\abstract
{
In a probabilistic framework of the interpretation of the initial mass function (IMF), the IMF cannot be arbitrarily normalized to the total mass, $\cal M$, or number of stars, $\cal N$, of the system. Hence, the inference of $\cal M$ and $\cal N$ when partial information about the studied system is available must be revised. (i.e., the contribution to the total quantity cannot be obtained by simple algebraic manipulations of the IMF).
}
{We study how to include constraints in the IMF to make inferences about different quantities characterizing stellar systems. It is expected that including any particular piece of information about a system would constrain the range of possible solutions. However, different pieces of information might be irrelevant depending on the quantity to be inferred. In this work we want to characterize the relevance of the priors in the possible inferences.}
{Assuming that the IMF is a probability distribution function, we derive the sampling
distributions of $\cal M$ and  $\cal N$ of the system constrained to different types of information 
available.}
{
We show that the value of $\cal M$ that would be inferred  must be described as a probability distribution $\Phi_{\cal M}[{\cal M};  m_\mathrm{a}, N_\mathrm{a},  \Phi_{\cal N}(\cal N)]$ that depends on the completeness limit of the data, $m_\mathrm{a}$, the number of stars observed down to this limit, $N_\mathrm{a}$, and the prior hypothesis made on the distribution of the total number of stars in clusters,  $\Phi_{\cal N}(\cal N)$.  
}
{}


\authorrunning{Cervi\~no et al.}
\titlerunning{The IMF and  the $\cal M$ inference}
\maketitle

\section{Introduction}

The study of cluster dynamics and star formation relies on the knowledge of cluster masses and the amount of such mass transformed into stars, $\cal M$. In most cases, we have partial information of the system, i.e., the observations of some  stars in the cluster. Such information is usually used in the inverse problem  using the initial mass function (IMF) {\it realization} (see below)
 as a distribution by number  to make inferences about a theoretical probability distribution function, the IMF $\phi(m)$  \citep{Bouetal98, Brietal02, Luhetal03, Olietal09, Bayoetal11}. However, such information is not enough to obtain cluster masses, and for some astrophysical studies it is required to assume a $\phi(m)$  covering all the range of possible stellar masses to make inferences about global cluster properties (the direct problem). 

This  use of the term IMF  for both the distribution by number for the inverse problem of statistics and the probability distribution function (pdf) for the direct problem can lead to different interpretations of the IMF itself and the results obtained from it \cite[cf.][ hereafter Paper I]{Ceretal12}. In this work, following \cite{Sca86}, we will adopt the pdf definition\footnote{This definition implies that stellar masses are identically and independent distributed, we refer Paper I for more details.}. 

The shape of the pdf and that of the distribution by number depend
crucially on the size of the sample, that is, the number of stars
${\cal N}$;  for large ${\cal N}$ values, the two shapes tend to be similar. However, this similarity can mislead one into believing that the distribution by number is just a scaled-up version of the pdf, with ${\cal N}$ being the scale factor. 
This would be very wrong since the physical meanings of both distributions are intrinsically different; Paper I is dedicated to exploring the consequences of this essential difference.

As a consequence, the standard methodology used to infer $\cal M$ values, which assumes the use of a correction factor 
for unobserved stars, is no longer valid.
The main goal of this paper is  to define a methodology based on the probabilistic approach of the IMF
 to obtain the total stellar mass ${\cal M}$ of an stellar sample from limited
information on the sample itself. 

This task is far from
trivial as we have to bridge different gaps
according to the amount of unknown information. We start the discussion by making an inventory of possible scenarios that differ from each other according to the amount of information available, with the aim of emphasizing how this affects the determination of $\cal M$ and $\cal N$.
Five such scenarios are:

\begin{enumerate}
\item We know (from the IMF) the probability of a random star  having a mass $m_\mathrm{star}$ equal to or larger than some given value $m_a$, but no specific information on the particular cluster is known.
 \label{item:imf}
\item We know (from observations) the number of stars ${\cal N}$
    in a particular cluster; we also know (from the IMF) the expected number of stars with $m\ge m_\mathrm{a}$.
    \label{item:prob}
\item We know (from observations) the number of stars ${\cal N}$
    in a particular cluster; we also know (from observations, too) that $N_\mathrm{a}$ stars have $m\ge m_\mathrm{a}$ and the mass of such stars.
\label{item:frequency}  
\item We know  that a particular cluster has $N_\mathrm{a}$ stars with $m\ge m_\mathrm{a}$ and the mass of such stars from observations.
\label{item:number}
\item We know  that a particular cluster has $N_\mathrm{a}$ stars with $m\ge m_\mathrm{a}$ and the mass of such stars, and we also know its total mass $\cal M$.
\label{item:mass}
\end{enumerate}

In scenario~\ref{item:imf}, which relies solely on knowledge of the IMF, we only know a theoretical probability 
that is independent of $\cal N$  and $\cal M$. Consecuently, we have neither information on $\cal M$ nor on the actual value of $m_\mathrm{star}$.

In scenario~\ref{item:prob}, we know that the cluster is the result
of sampling the IMF with ${\cal N}$ stars. With such information, we can compute the sampling distribution of $\cal M$:
that is, the distribution of possible values of $\cal M$ constrained by the value of $\cal N$.
In particular, if ${\cal N} = 1$ the distribution of total masses is the IMF itself, and if ${\cal N} \rightarrow \infty$, 
the distribution of $\cal M$ is a Gaussian, because of the central limit theorem.
In all intermediate cases, the  sampling distribution of $\cal M$ at a given ${\cal N}$ is a more or less asymmetric function, which in turn implies that its mean 
value $\left< {\cal M} \right>$  is not (in general) the same as its most 
probable value.

Scenario~\ref{item:frequency}
is a constrained version of the 
previous one. In the universe of all possible clusters with $\cal N$ stars, only those {\it conditioned to} have $N_\mathrm{a}$ stars with 
mass equal to or larger than $m_\mathrm{a}$ can represent the cluster studied. The resulting distribution of possible $\cal M$, which is different from the previous sampling distribution, can be obtained by imposing an a posteriori condition on it. However, since we also know the mass of the $N_\mathrm{a}$ stars, an additional constraint must be applied

Scenario~\ref{item:number}  only constrains  $\cal M$ to be equal to or larger than the contribution of the $N_\mathrm{a}$ stars.
We cannot progress further unless we additionally assume a distribution of possible ${\cal N}$ values. 
If we do so, the resulting values of the mean total mass $\left< {\cal M} \right>$ and the most probable value will differ from
those obtained under scenario~\ref{item:prob}, since in the present case $\cal N$ is not fixed but distributed
and this  affects the shape of the sampling distribution of  $\cal M$.

In scenario~\ref{item:mass}, we know that the mass is ${\cal M}$ and that there are $N_\mathrm{a}$ stars 
with $m \ge m_\mathrm{a}$. 
The probability 
distributions that describe such a cluster (such as, for example, the
distribution of possible $\cal N$ values or of the $N_\mathrm{a}$ most massive stars
that the cluster could host) correspond to the 
particular situation described in scenario~\ref{item:number} with the additional constraint of knowing
${\cal M}$. 

From the above discussion, it is clear that the $\cal M$ derived in each of the above scenarios are different.
Although all the resulting distributions are
derived from the IMF, each of them is the result of including different pieces of information in the analysis: either the total number 
of stars  $\cal N$ in the cluster (in scenarios~\ref{item:prob} and \ref{item:frequency}), and its probability 
distribution (in scenarios~\ref{item:number} and \ref{item:mass}) or the presence 
of $N_\mathrm{a}$ stars above a given mass value  (in scenarios~\ref{item:frequency},
 \ref{item:number}, and \ref{item:mass}). Each case results in a different conditional probability distribution, 
 which results in a different estimation of $\cal M$.

We note that relating the IMF with the corresponding sampling (and 
conditional) probability distributions is correct, given the set under study. We also note that we have an additional piece of information in such a set: stars are individual, discrete entities (i.e., $\cal N$ is a natural number). Such a condition must be fulfilled by any cluster in the Universe and must be included in all scenarios as a restriction (even in cases where there is no explicit reference to $\cal N$, as in scenario~\ref{item:mass}).

The preceding discussion boils down to the following point: as an underlying density distribution, 
the IMF  describes neither a particular case nor any observational constraints (such as, e.g., 
the number of stars with a given mass observed in a particular cluster). 
Once an observational constraint is included (e.g., the fact that one star with known mass is present), conditional probabilities must be applied.
Stated otherwise, the distribution that describes the universe of possible results (the IMF) is an  a priori probability, 
and the probability constrained to the observed data is an a posteriori (conditional) probability. 
Confusing the  a posteriori 
probability with the a priori probability is one of the most common flaws in hypothesis testing reasoning \cite[this is also called the {\it Prosecutor's fallacy}: see ][~for a 
discussion in a similar astrophysical context]{SM08}. In these situations, it is fundamental understand the true context of the question before  seeking an answer. This has been done in the five scenarios discussed above.

The structure of the paper is as follows: In Sect.~\ref{sec:formal} we summarize the basic concepts required
to use the IMF in a probabilistic framework (see Paper I for a more extended discussion).
In Sect.~\ref{sec:totalmass1} we consider an ideal case in which all
the stars in the system are known. Then we replace known information
by unknowns to describe real situations where the use of the
IMF or a related sampling distribution is required. Section~\ref{sec:astro} 
shows the methodology to obtain $\cal M$ from partial information of the system in the scenarios presented above and their application to some 
astrophysical cases. We discuss some considerations about the use of prior information in Sect.~\ref{sec:discusion}.
Our conclusions are described in Sect.~\ref{sec:conclusions}.

\section{Formal probabilistic formulation}
\label{sec:formal}

The basis of the probabilistic formulation has been presented in Paper I. We refer to that paper for more details and  include here only the 
basic formulae needed for this work.

\begin{enumerate}
\item The IMF, $\phi(m)=\mathrm{d}N/\mathrm{d}m$, is a probability density function (pdf), which can be integrated over a given mass range to derive the 
probability of finding a star in that range. The mass limits $\mathrm{m_{low}}$ and $\mathrm{m_{up}}$ are given by
stellar theory and must fulfill $\int_{\mathrm{m_{low}}}^{\mathrm{m_{up}}} \phi(m) \mathrm{d}m = 1$; that is, we are certain that any possible star has a mass between $\mathrm{m_{low}}$ and $\mathrm{m_{up}}$.

The probability of a random star having a mass {\it lower than} a given value $m_\mathrm{a}$ is given by

\begin{equation}
p(m < m_\mathrm{a}) =  \int_{\mathrm{m_{low}}}^{m_\mathrm{a}} \phi(m) \, \mathrm{d}m.
\label{eq:pltm}
\end{equation}

In this work, the integrals over the IMF will always be
read as {\it equal to or larger than} the lower limit and  {\it lower than} the upper limit.

In this work we employ the Kroupa IMF \citep{Kro01,Kro02} as 
used  in \cite{WK06}, with $\mathrm{m_{up}} = 120 \mathrm{M}_\odot$, 
$\mathrm{m_{low}} = 0.01 \mathrm{M}_\odot$, and a correction of $k'=1/3$ for stars
with mass lower than 0.08 $\mathrm{M}_\odot$\footnote{Such a correction was not used in
Paper I. However, it is the parametrization used in the set of clusters by \cite{KM11} we  use in this work
for comparing methodologies.}

\item Different observational scenarios can be described by adding constraints to the IMF.  
For instance, we may explicitly include a limit on $m\mathrm{_{a}}$ and  compute probabilities for stars with masses lower than  $m\mathrm{_{a}}$.  In this case, we must define an a posteriori pdf related to the IMF that includes such a condition:

\begin{eqnarray}
\phi(m | m < m\mathrm{_{a}}) = 
\frac{\phi(m) \, \mathrm{H}(m\mathrm{_{a}}-m)}{p(m <  m\mathrm{_{a}})},
\label{eq:IMFmcond}
\end{eqnarray}

\noindent where $\mathrm{H}(m_\mathrm{a} - m)$ is the Heaviside function\footnote{We use here the Heaviside function as a distribution to define the domain of $\phi(m)$ including constraints. In this situation the value of $\mathrm{H}(0)$ is {\it not} defined, but it is assigned  a posteriori to be consistent with the convention used in the integral limits. In the case of Eq.~\ref{eq:IMFmcond}, $\mathrm{H}(0)=0$.}, which ensures that no star equal to or larger than  $m_\mathrm{a}$ can be present in the cluster. 
We note that $\phi(m | m < m\mathrm{_{a}}) $ is a pdf also. The mean mass of such distribution is

\begin{equation}
 \left< m  | m < m\mathrm{_{a}} \right> = \frac{ \int_{\mathrm{m_{low}}}^{\mathrm{m_{up}}} \, m \, \phi(m) \, \mathrm{H}(m\mathrm{_{a}}-m) \mathrm{d}m}{p(m < m\mathrm{_{a}})}.
 \label{eq:mmeancond}
\end{equation}

\item The pdf describing ensembles with a total number of stars $\cal N$ 
(formally,  a sampling distribution conditioned to have $\cal N$ stars) can be calculated as successive convolutions 
of the corresponding pdf for one star. 
For instance, the pdf  for the total mass, $\Phi_{\cal M}({\cal M}|{\cal N})$,  is 
the result of convolving the IMF ${\cal N}$ times in a recursive convolution 
\cite[see][]{CLCL06,SM08}:

\begin{equation}
\Phi_{{\cal M}}({\cal M}|{\cal N}) = \overbrace{ \phi(m)\otimes \phi(m) \otimes \, .... \,\otimes 
\phi(m)}^{{\cal N}}.
\label{eq:Mtot}
\end{equation}

The same procedure applies to any other pdf. The mean value of the resulting distribution is

\begin{equation}
 \left< {\cal M} | {\cal N} \right> = {\cal N} \times \left< m \right> = {\cal N} \times \int_{\mathrm{m_{low}}}^{\mathrm{m_{up}}} \, m \, \phi(m) \, \mathrm{d}m.
 \label{eq:meanvaluegen}
 \end{equation}
 
Mean values of constrained distributions when sampled with $\cal N$ stars are obtained in a similar way.

\end{enumerate} 

\section{The trade-off between knowledge and probability}
\label{sec:totalmass1}

Once we have laid down the basic framework, we apply it to our science case: the estimation of the total mass $\cal M$ of a cluster from
a partial knowledge of its stellar content. To do that we progressively replace known information
by unknowns to describe real situations;  however, the following items here are not directly related to the scenarios quoted in the Introduction (we will come back to such scenarios in Sect.~\ref{sec:astro}).

\subsection{Case study 1: Everything is known}
\label{subsec:caso1}

We begin with an ideal observational point of view, where we suppose that we know the masses 
$m_i^{\mathrm{obs}}$ of  every one of the ${\cal N}$ stars in a cluster. Thus, the total mass,
${\cal M}$, is also known. 
In this hypothetical case, it is not required to use the IMF. However, this exercise allows us to illustrate the trade-off 
between the use of known data from a particular cluster (i.e., a particular IMF realization) and
 the use of probability distributions.

We sort the stars in ascending order according to their mass. We use a subindex in brackets to denote that such 
operation has been performed, so $m_i$ is the $i$-th random sampled element and $m_{[i]}$ is the $i$-th element after
sorting the data.  We also assume that  the most massive 
star has a mass  $m_{[{\cal N}]}^{\mathrm{obs}} = m_\mathrm{max}^{\mathrm{obs}}$ with a value 
lower than $\mathrm{m_{up}}$. 

In addition, we assume that we have $N_{\mathrm{a}}$ stars equal or more massive than an arbitrary value 
$m_\mathrm{a}$, so that $m_{[{\cal N}-N_\mathrm{a} ]} <  m_\mathrm{a}$, and $m_{[{\cal N}-N_\mathrm{a} +1]}\ge m_\mathrm{a}$. We express the total number of stars and total mass as a function  of the  
$N_{\mathrm{a}}$ set. It  can be 
described as

\begin{equation}
N_\mathrm{a} = \sum_{i={\cal N}-N_\mathrm{a}+1}^{\cal N}  \delta_{i,i}, ~~~~~ M_\mathrm{a} = \sum_{i={\cal N}-N_\mathrm{a}+1}^{\cal N}  m_{[i]}^\mathrm{obs} \delta_{i,i},
\end{equation}

\noindent  where  $\delta_{i,j}$ is the Kronecker delta.  
The  total mass in the ensemble is

\begin{equation}
{\cal M}  = M_\mathrm{a}   + \sum_{i=1}^{{\cal N}-N_\mathrm{a}} m_{[i]}^{\mathrm{obs}} \delta_{i,i}.
\label{eq:MtotKnown}
\end{equation}

\noindent  These two equations, rewritten in terms of frequencies and mean stellar mass {\it in the complete sample}, are, respectively

\begin{eqnarray}
\frac{N_\mathrm{a}}{\cal N} &=& \sum_{i={\cal N}-N_\mathrm{a}+1}^{\cal N}  \frac{\delta_{i,i}}{{\cal N}},
\label{eq:mmax_sum} \label{eq:frecNa}
\end{eqnarray}

\noindent and

\begin{eqnarray}
\left< \widetilde{m} \right> &=&   \frac{N_\mathrm{a}}{{\cal N}}  \,\,  \frac{M_\mathrm{a}}{N_\mathrm{a}}  
 + \frac{{\cal N}-N_\mathrm{a}}{{\cal N}} \sum_{i=1}^{{\cal N}-N_\mathrm{a}} \frac{m_{[i]}^{\mathrm{obs}}}{{\cal N}-N_\mathrm{a}}. \label{eq:frecM}
\end{eqnarray} 

Multiplying $\left< \widetilde{m} \right>$ by $\cal N$ produces the value of $\cal M$. However, we note that conceptually

\begin{eqnarray}
{\cal M} = {\cal N}\, \times \,  \left< \widetilde{m} \right>  \neq {\cal N}\, \times \,  \left< m \right> = \left< {\cal M} \right>,
\end{eqnarray}

\noindent since $\left< \widetilde{m} \right>$ (the sample mean) does not coincide with the the mean stellar mass 
obtained from the IMF, $\left< m \right>$ (the population mean). That is, $\left< \widetilde{m} \right>$ is an estimate of $\left< m \right>$  obtained from a sample of ${\cal N}$ stars, so, formally, $\left< \widetilde{m} \right> = \left< \widetilde{m}  | {\cal N}\right>$. In the following, we use the $\widetilde{m}$ symbol to denote an estimate of $m$. In the computation of this estimate, the value of ${\cal N}$ must be taken into consideration, although we will not write it explicitly in order to simplify the notation.

\subsection{Case study 2: The total number of stars and the mass of the most massive $N_\mathrm{a}$ stars are known}
\label{subsec:NaNKnow}

In this case we have less information than in the previous case since we only know 
 $m_{[i]}^{\mathrm{obs}}$ with $i=\{{\cal N}-N_\mathrm{a}+1,\dots\,{\cal N}\}$, stellar masses, and ${\cal N}$.
But we had seen that estimates obtained from actual values, such as $\left< \widetilde{m} \right>$
can be related to values obtained from the IMF. So we can replace these estimates with

\begin{eqnarray}
\sum_{i=1}^{{\cal N}-N_\mathrm{a}} \frac{m_i}{{\cal N}-N_\mathrm{a}} = \left< \widetilde{m} | m < m\mathrm{_{a}^{obs}} \right> \rightarrow  \left< m | m < m\mathrm{_{a}^{obs}} \right>. \nonumber
\end{eqnarray}

Thus, although we cannot know the actual $\cal M$ value, we can at least  obtain average values given different sets of constraints:

\begin{eqnarray}
\left<{\cal M} |  m_{[i]}^\mathrm{obs} \ge m_\mathrm{a}^\mathrm{obs}\; i={\cal{N}} - N_\mathrm{a} +1, ... \cal{N};~{\cal N}  \right> =   \nonumber\\
 \,\,\,\,\,\,\,\,\,\,\,\,\,\,\,\,\,\,\,\,\,\,\,\,\,\,\,M_\mathrm{a}
 + ({\cal N}-N_\mathrm{a})  \left< m | m < m\mathrm{_{a}^{obs}} \right>. \label{eq:frecprobM}
\end{eqnarray}

This illustrates the trade-off between observed frequency distributions and probability: when we use a probability distribution, we cannot have access to the actual values, but  we can have access  to the distribution of {\it possible} values and the mean value of {\it all} these possible values. 
In this case we are using the estimates argument in the opposite direction to a statistical analysis, i.e., we are making the assumption that  {\it all} the stars are distributed following the IMF\footnote{Hence, it includes the $N_\mathrm{a}$ subset  with known stellar masses.} and using it to make inferences about related quantities.

\subsection{Case study 3: Only the mass of the $N_\mathrm{a}$ more massive stars is known}
\label{subsec:NaKnow}

Observations of clusters in many cases only allow characterization of the $N_\mathrm{a}$ most luminous stars with masses  $m_{[i]}^\mathrm{obs}, \, i=\{{\cal N}-N_\mathrm{a}+1,\dots\,{\cal N}\}$. They also lack a proper census that includes the lowest luminous members \cite[see][ as counterexamples]{Bayoetal11,KM11}. In this case, it is more difficult to obtain estimates, since we can not define a {\it frequency} of $N_\mathrm{a}$. Therefore, is the following reasoning valid?

\begin{eqnarray}
 \widetilde{p}(m | m < m\mathrm{_{a}^{obs}}) = \frac{{\cal N} - N_\mathrm{a}}{\cal N} \rightarrow p(m | m < m\mathrm{_{a}^{obs}}), \nonumber \\
 \widetilde{p}(m | m \geq m\mathrm{_{a}^{obs}}) = \frac{N_\mathrm{a}}{\cal N} \rightarrow p(m | m \geq m\mathrm{_{a}^{obs}}). \nonumber 
\end{eqnarray}

\subsubsection{When is the correspondence ${\cal N} = N_\mathrm{a} / p(m | m \geq m\mathrm{_a})$ valid?}

We  divide the IMF in, e.g., $k+1$ mass intervals, where the mass interval containing the lower masses, e.g., the $k+1$, comprises the ${\cal N}Ê- N_\mathrm{a}$ of unknown stars with mass lower than $m_\mathrm{a}$. Each of the remaining $i$ mass interval, which belong to $[m_i^\mathrm{low}, m_i^{\mathrm{up}})$ contain $n_i$ stars\footnote{In this case, we are distributing the known $N_\mathrm{a}$ stars in $k$ {\it intervals} and  not using the particular $m$ values of known stars. Such intervals can be arbitrary and must only obey the condition $\sum_{i=1}^{k} n_i = N_\mathrm{a}$. So the index $i$ here refers to the interval, not to a particular stellar mass.},
 so that $\sum_{i=1}^{k} n_i = N_\mathrm{a}$. The probability of having a star in a given mass interval is given by the integration of the IMF over such a mass interval, $p_i(m) = p(m \in [m_i^\mathrm{low}, m_i^{\mathrm{up}})$). 
We assume that the cluster is a random realization of the IMF for $\cal N$ stars, so  the probability of having the $\cal N$ stars distributed in the $k+1$ intervals with $n_i$ stars in the $i$-th interval for a given (unknown) number of stars $\cal N$ is given by the multinomial distribution\footnote{Since $\cal N$ is a discrete quantity, their pdf directly provides the probability. In addition, the distribution can be also expressed as a binomial distribution with $A(p_i,n_i) \, N_\mathrm{a}! = p(m \geq m_\mathrm{a})^{N_\mathrm{a}}=(1-p(m < m_\mathrm{a}))^{N_\mathrm{a}}$.} 

\begin{eqnarray}
\Phi_{N_\mathrm{a}}(N_\mathrm{a}|{\cal N}) &=& {\cal P}(m \ge m_\mathrm{a}, \sum_{i=1}^{k} n_i = N_\mathrm{a}|{\cal N})  = \nonumber\\
&=& \frac{{\cal N}!}{({\cal N} - N_\mathrm{a})! \prod_{i=1}^{k} n_i!} p(m < m_\mathrm{a})^{{\cal N} - N_\mathrm{a}} \, \prod_{i=1}^{k}p_i(m)^{n_i} \nonumber \\
&=& A(p_i,n_i) \; \frac{{\cal N}!}{({\cal N} - N_\mathrm{a})!} p(m < m_\mathrm{a})^{{\cal N} - N_\mathrm{a}},
\label{eq:multinom}
\end{eqnarray}

\noindent where we have included in $A(p_i,n_i)$ all the known information. However, we are interested in the complementary distribution $\Phi_{\cal N}({\cal N} | N_\mathrm{a})$, which must be obtained using the Bayes' theorem (see, e.g., Paper I). Assuming that the possible values of $\cal N$, $\Phi_{\cal N}({\cal N})$ follow
a discrete power-law probability distribution with exponent $-\beta$, we obtain

\begin{equation} 
\Phi_{\cal N}({\cal N} | N_\mathrm{a}) =  A'\, \frac{{\cal N}!}{({\cal N} - N_\mathrm{a})!} \,  p(m < m_\mathrm{a})^{{\cal N} -N_\mathrm{a}}  \, {\cal N}^{-\beta}, 
\label{eq:phiNNa}
\end{equation}

\noindent where $A'$ is a normalization value that includes all the known terms and is independent of $A(p_i,n_i)$ since $A(p_i,n_i)$  is canceled out by the normalization constant, Thus,  the inference about the total number of stars only depends on the number of stars $N_\mathrm{a}$ more massive than a certain  observational value $m_\mathrm{a}$, and not on the particular distribution of such stars in different mass bins. 

This result might seem surprising: the knowledge of the masses of particular stars does not provide additional information on (the number) of unobserved ones\footnote{However, we note that such information is still relevant for the computation of $\cal M$: the individual masses of stars more massive than $m_\mathrm{a}$ provide the amount of mass in the mass range, 
$M_\mathrm{a}$.}. It can be argued that, for example, an excess or deficit of the observed number of stars in a given mass range constrains the total number of stars from being compatible with sampling effects. However, such arguments are valid for IMF inferences (which IMF shape is more probable, given some observations?), i.e., the problem of obtaining the IMF.  

In our case, {\it a given IMF is assumed and the observations are a random realization of it}. The particular observed set may be a highly improbable (but still possible) realization of the assumed IMF. Nevertheless, whatever its a priori probability of happening, {\it it has actually happened}, and thus a posteriori probabilities must be obtained by taking this fact into consideration. In addition, since stellar masses are random variables (cf. Paper I), the occurrence of having a star (or a set of stars) with a given particular mass has no impact on the individual masses of the remaining stars.

The mode of  $\Phi_{\cal N}({\cal N} | N_\mathrm{a})$,  ${\cal N}^\mathrm{mode}$ 
is obtained by equating to zero its first derivative with respect to ${\cal N}$, which, for large $\cal N$ values\footnote{\label{footnote:NaMt}In practical terms it implies large $N_\mathrm{a}$ values. Actually, $\Phi_{\cal N}({\cal N} | N_\mathrm{a})$ is a discrete distribution, hence not derivable, but the formulae provide a reasonable value as far as the Stirling approximation of factorial functions are valid, i.e., $\cal N$, $N_\mathrm{a}$, and ${\cal N} - N_\mathrm{a}$ larger than 15.} , yields

\begin{equation}
{\cal N}^\mathrm{mode} \approx \frac{\beta-N_\mathrm{a}}{\ln  p(m < m_\mathrm{a})},
\label{eq:Nmode}
\end{equation}

\noindent where we used the Stirling approximation of factorial functions and a first-order Taylor approximation of logarithm functions valid for $\beta \ne 0$. In the case of a flat distribution with $\beta =0$, the approximate mode of the distribution is obtained by solving

\begin{equation}
 p(m < m_\mathrm{a}) = \left( 1 - \frac{N_\mathrm{a}}{{\cal N}^\mathrm{mode}}\right),
\end{equation}

\noindent which coincides with the estimation of the probability $\widetilde{p}(m < m_\mathrm{a})$ for known $N_\mathrm{a}$ and $\cal N$.
This means that $N_\mathrm{a} / p(m | m \geq m\mathrm{_{a}})$ provides the mode ${\cal N} ^\mathrm{mode}$ of $\Phi_{\cal N}({\cal N} | N_\mathrm{a})$ assuming a flat $\Phi_{\cal N}({\cal N})$ distribution. However, we know that the initial cluster mass function (ICMF, $\Phi_{\cal M}({\cal M})$ ) is not flat \citep{LL03,Pisetal08} and that it must be somehow related to $\Phi_{\cal N}({\cal N})$ (cf. Eq.~\ref{eq:Mtot}),
although we are not able to establish its functional form. Whatever equation we use to obtain ${\cal N} ^\mathrm{mode}$, we are left in  the uncomfortable situation of mixing a mean value ($ \left< m | m < m\mathrm{_{a}} \right> $) with a mode value ${\cal N} ^\mathrm{mode}$ to obtain an inference about $\cal M$. However,  we have no means to give a meaning of this inference: Is it a mean, a mode, on any other parameter?

This suggests that it is better to use the resulting probability distribution of ${\cal N}(N_\mathrm{a})$ and obtain the corresponding  $\Phi_{\cal M}[{\cal M} | N_\mathrm{a}, \Phi_{\cal N}({\cal N})]$ to make inferences about $\cal M$. In addition, this way to proceed is in agreement with the International Organization for Standardization 
(ISO), which recommends expressing  the uncertainty in the results as a pdf\footnote{{\it Guide to the Expression of Uncertainty in Measurement} 
(International Organization for Standardization, Switzerland, 1995) {\tt http://www.bipm.org/en/publications/guides/gum.html}.}.

\section{Use cases}
\label{sec:astro}

\begin{table}
\begin{tabular}{l|rrrrrr}
\hline
name & $\cal M$ & $M_\mathrm{a}$ & $\cal N$  & $N_\mathrm{a}$ & $\left<\tilde{m_\mathrm{a}}\right>$ & $\log p_\mathrm{nor}(N_\mathrm{a}|{\cal N})$\\
\hline
Tau. & &&&&&\\
\#1& 10.6&   7.6&  20&   8&  0.95& -1.06\\
\#2& 15.5&  11.5&  30&  12&  0.96& -1.55\\
\#3&   8.1&   5.9&  19&   8&  0.74& -1.20\\
\#4& 22.7&  21.7&  24&  18&  1.20& -7.56\\
\#5&   8.2&   8.0&  14&  10&  0.80& -3.91\\
\#6& 17.7&  14.7&  31&  14&  1.05& -2.39\\
\#7& 16.1&  13.9&  24&  13&  1.07& -3.20\\
\#8&  12.3&  10.2&  16&   5&  2.05& -0.34\\
(field)& 88.5&  72.3& 174&  73&  0.99&-10.15\\
\hline
ChaI & &&&&&\\
\#1&    3.7&   1.7&  12&   2&  0.85&  0.00\\
\#2&  40.5&  25.9&  96&  20&  1.30& -0.04\\
\#3&  21.7&  16.4&  43&  16&  1.03& -1.69\\
(field)&  42.6&  30.1&  86&  27&  1.11& -1.58\\
\hline
Lup.3 & &&&&&\\
\#1&   18.2&  13.4&  36&  11&  1.22& -0.62\\
(field)&  18.1&  12.9&  34&  11&  1.17& -0.76\\
\hline
IC348 & &&&&&\\
\#1& 111.9&  87.6& 186&  65&  1.35& -5.43\\
\#2&    3.1&   0.5&  11&   1&  0.53& -0.08\\
(field)&  78.2&  51.7& 166&  35&  1.48& -0.08\\
\hline
\end{tabular}
\caption[]{Data form stellar associations by \cite{KM11}. We show the total mass ($\cal M$), the mass into stars more massive than $m_\mathrm{a}$ ($M_\mathrm{a}$), the total number of stars ($\cal N$), the number of stars more massive than $m_\mathrm{a}$ ($N_\mathrm{a}$), the estimation of the mean mass for stars more massive than $m_\mathrm{a}$ ($\left<\tilde{m_\mathrm{a}}\right> = \left<\tilde{m}| m \geq m_\mathrm{a}\right>$), and the logarithm of the probability that a cluster with $\cal N$ stars following the assumed IMF would have $N_\mathrm{a}$ stars with mass equal or larger than $m_\mathrm{a}$ divided by the maximum of such distribution
 ($\log p_\mathrm{nor}(N_\mathrm{a}|{\cal N})$). The $m_\mathrm{a}$ value is set to $0.5~\mathrm{M}_\odot$.}
\label{tab:table1}
\end{table}

Having presented the probabilistic framework and the related information trade-off, we can compare the probabilistic methodology and the distribution by number methodology to obtain $\cal M$ and $\cal N$. For comparison purposes, we have used the data from \cite{KM11} to illustrate the differences. The data contain the observed masses for individual stars belonging to 14 young stellar groups in four different regions. They also contain the stellar mass of field stars in the four analyzed regions.
Table \ref{tab:table1}Ê~shows the identifier of the cluster along with the values of $\cal M$, $M_\mathrm{a}$, $\cal N$, $N_\mathrm{a}$, and the estimation of the mean mass, $\left<\tilde{m_\mathrm{a}}\right>= \left<\tilde{m}| m \geq m_\mathrm{a}\right>$ 
from the census of stars with $m \geq m_\mathrm{a}$.
\cite{KM11} state that their mass estimates are valid with a relative error of 50\%; in this work we assume that the tabulated values can be taken at face value without errors. 
They also state that their census is complete at a 90\% level down to 0.08 $\mathrm{M}_\odot$; hence their total mass estimation would be actually a lower limit of the real value. Whatever the case, we assume again that the $\cal M$ values obtained from the data can be use at face value without errors. Finally, we assume  that the data is complete at 100\% down to $m_\mathrm{a} = 0.5 \mathrm{M}_\odot$.
We use this $m_\mathrm{a}$ value to illustrate the $\cal M$ inference in scenarios 2, 3, and 4 in the introduction.

As reference, the IMF used here produces $\left<m \right> = 0.46 \mathrm{M}_\odot$,  $\left<m| m \geq m_\mathrm{a}\right> = 1.64 \mathrm{M}_\odot$, and  $p(m | m \geq m\mathrm{_{a}}) = 0.19$. We can make a first-order test about the compatibility of the cluster data 
with the assumed IMF by computing the probability of having a given $N_\mathrm{a}$ number of stars with mass larger than $m_\mathrm{a}$ in a cluster with $\cal N$ stars. It can be done by dividing  the IMF into two bins, $[\mathrm{m_{low}}, m_\mathrm{a})$ and $[m_\mathrm{a},\mathrm{m_{up}})$, and using the probability in each bin to define a binomial distribution. The logarithm value of the resulting probabilities normalized to the maximum value of the distribution,  $\log p_\mathrm{nor}(N_\mathrm{a}|{\cal N})$, are shown in column 7 of Table~\ref{tab:table1}\footnote{We note that a comparison of $\tilde{p}(m|m\geq m_\mathrm{a})$ and $p(m|m\geq m_\mathrm{a})$ does not produce a valid test about IMF compatibility, since the importance of the possible deviations depends on how many stars are in the sample (size of sample effects). Interestingly, IC348 \#1, which deviates from the IMF in this test, is the system used as an example by \cite{KM11} to argue that their systems follows a Kroupa IMF (their Fig. 6). Although the shape of the IMF realization in IC348\#1 would look like a Korupa IMF, the deviations (fluctuations) observed are actually too large compared with the expected ones taking into account the number of stars in the system.}. In this test we see that our hypothesis about the validity of the used IMF in all the associations is actually questionable
for the stars in Taurus field, Taurus \#4, and IC348 \#1, and  would produce some problems in the analysis of Taurus \#5, \#7, and \#6.

\subsection{Distribution-by-number methodology}

The distribution-by-number methodology considers that the IMF can be used with an arbitrary normalization. Such normalization can be either to $\cal N$ or $\cal M$, which implies multipling $\phi(m)$ by $\cal N$ or ${\cal M}/\left<m\right>$, respectively. In addition, it is assumed that $\cal N$ and $\cal M$ are deterministically related by the relation

\begin{equation}
{\cal M} = {\cal N} \times \left< m \right>.
\label{eq:Mdet}
\end{equation}

This provides $\cal M$ in all the cases where $\cal N$ is given and vice versa. We can include additional information like $M_\mathrm{a}$ and $N_\mathrm{a}$ to make alternative inferences about $\cal M$. Following the procedure of this paper, the most information is included using a formula similar to Eq.~\ref{eq:frecprobM}:

\begin{equation}
{\cal M} = M_\mathrm{a}   + ({\cal N}-N_\mathrm{a})  \left< m | m < m\mathrm{_{a}} \right>.
\label{eq:MdetfromNMyNtot}
\end{equation} 

However, we can choose to use only partial information, such as the contribution of $M_\mathrm{a}$ to the total budget. Then the ratio ${\cal M}/M_\mathrm{a}$ is constant, and is equal to the ratio of $m \timesÊ\phi(m)$ integrated in the whole range, $\left< m \right>$, over the same function integrated in the $m_\mathrm{a}$, $\mathrm{m_{up}}$ range. As a result, 
 $\cal M$ is:

\begin{equation}
{\cal M} =  \frac{M_\mathrm{a} \times \left< m \right>}{\int_{m_\mathrm{a}}^\mathrm{m_{up}} m\,\phi(m)\,\mathrm{d}m}.
\label{eq:MdetformM}
\end{equation}

On the other hand, we could choose to use the contribution of  $N_\mathrm{a}$ to the total budget. Then the ratio ${\cal N}/N_\mathrm{a}$ is constant 
 and is equal to the ratio of $\phi(m)$ integrated in the whole range (that is, the unity) over the $\phi(m)$ integrated in the $m_\mathrm{a}$, $\mathrm{m_{up}}$ range. Since ${\cal M} = {\cal N} \times \left< m \right>$, 
 $\cal M$ is

\begin{equation}
{\cal M} =  \frac{N_\mathrm{a} \times \left< m \right>}{\int_{m_\mathrm{a}}^\mathrm{m_{up}} \phi(m)\,\mathrm{d}m}.
\label{eq:MdetformN}
\end{equation}

We could also choose to use just $M_\mathrm{a}$ and $N_\mathrm{a}$ values without the information about $\cal N$ (similar to Eq.~\ref{eq:MdetfromNMyNtot} with some additional algebraic manipulation):

\begin{equation}
{\cal M} = M_\mathrm{a}  + N_\mathrm{a} \,  \left< m | m < m\mathrm{_{a}} \right> \, \frac{\int_{\mathrm{m_{low}}}^{m_\mathrm{a}} \phi(m)\,\mathrm{d}m}{\int_{m_\mathrm{a}}^\mathrm{m_{up}} \phi(m)\,\mathrm{d}m}.
\label{eq:MdetformNyM}
\end{equation}

Equations \ref{eq:MdetformM}, \ref{eq:MdetformN}, and \ref{eq:MdetformNyM}  produce an equal value of $\cal M$ as far as

\begin{eqnarray}
\frac{M_\mathrm{a}}{N_\mathrm{a}} = \left<\tilde{m}| m \geq m_\mathrm{a}\right>  \rightarrowÊ\left<m| m \geq m_\mathrm{a}\right>,\nonumber
\end{eqnarray}

\noindent and they will produce a result similar to Eq.~\ref{eq:Mdet} and \ref{eq:MdetfromNMyNtot} as far as, additionally, 

\begin{eqnarray}
 \widetilde{p}(m | m \geq m\mathrm{_{a}}) = \frac{N_\mathrm{a}}{\cal N} \rightarrow p(m | m \geq m\mathrm{_{a}}). \nonumber 
\end{eqnarray}

In relation to the scenarios presented in the Introduction, scenario 2 (only $\cal N$ is observed) is described by Eq.~\ref{eq:Mdet}. Scenario 3 ($\cal N$, $N_\mathrm{a}$, and $M_\mathrm{a}$ are known) can be described by Eqs.~\ref{eq:Mdet}, \ref{eq:MdetfromNMyNtot}, \ref{eq:MdetformM}, \ref{eq:MdetformN}, and \ref{eq:MdetformNyM}, depending the information we choose to use, with Eq.~\ref{eq:MdetfromNMyNtot} being the one that uses the most available information. Finally, scenario 4 can be described by Eqs.~
\ref{eq:MdetformM}, \ref{eq:MdetformN}, and \ref{eq:MdetformNyM}, with Eq.~\ref{eq:MdetformNyM} being the one that use the most available information.

The resulting $\cal M$ estimations from \cite{KM11} data employing this methodology are shown in Table~\ref{tab:table2}, which uses different information from the cluster. The inferred $\cal M$ varies depending on the formulae (and hence the amount of not redundant information) used for the inference. 
The best result is obtained by Eq.~\ref{eq:MdetfromNMyNtot}, but unfortunately it does not have a practical application ($\cal N$ is unknown most of the times). 

With respect to the equations that can be used in scenario 4 (the common observational case), Eq.~\ref{eq:MdetformNyM} produce a value between the results of Eqs.~\ref{eq:MdetformM} and \ref{eq:MdetformN}. Also, since $\left<\tilde{m}| m \geq m_\mathrm{a}\right>$ underestimates $\left<m| m \geq m_\mathrm{a}\right>$ for the clusters in the given sample, Eq.~\ref{eq:MdetformM} produces lower values than Eq.~\ref{eq:MdetformN} (see Taurus \#8 as the opposite example). The range of inferred $\cal M$ values covered by  Eq.~\ref{eq:MdetformM}, ~\ref{eq:MdetformN} and \ref{eq:MdetformNyM} only include the observed $\cal M$ value in four cases (Taurus \#8, Cha \#1 and \#2, and the field stars in IC348), suggesting a 20\% rate of success (33\% if we exclude the five clusters with possible strong deviations from the assumed IMF). In addition, we do not known which equation produces the more reasonable value (although Eq.~\ref{eq:MdetformNyM} is preferred) nor do we have a possible evaluation accuracy associated to each case.

\begin{table}
\begin{tabular}{l|r|r|rrr|r|}
 \hline
 & \multicolumn{5}{c|}{$\cal M$ inferred} & \\
 & Sce.2 & Sce.3 &\multicolumn{3}{c|}{Sce.4} & \\
name & Eq.~\ref{eq:Mdet} & Eq.~\ref{eq:MdetfromNMyNtot} & Eq.~\ref{eq:MdetformM} & Eq.~\ref{eq:MdetformN} &  Eq.~\ref{eq:MdetformNyM}  & ${\cal M}$ obs. \\
 \hline
Tau. & &&&&&\\
\#1&  9.2&   9.7&  11.0&  19.0&  13.4&  10.6\\
\#2& 13.8&  14.7&  16.7&  28.6&  20.3&  15.5\\
\#3&   8.8&   7.8&   8.5&  19.0&  11.8&   8.1\\
\#4& 11.1&  22.7&  31.3&  42.8&  34.9&  22.7\\
\#5&   6.5&   8.7&  11.5&  23.8&  15.3&   8.2\\
\#6& 14.3&  17.7&  21.2&  33.3&  24.9&  17.7\\
\#7&   11.1&  15.9&  20.1&  30.9&  23.5&  16.1\\
\#8&    7.4&  12.2&  14.8&  11.9&  13.9&  12.3\\
field& 80.3&  90.1& 104.5& 173.7& 125.8&  88.5\\
\hline
ChaI & &&&&&\\
\#1&   5.5&   3.5&   2.5&   4.8&   3.2&   3.7\\
\#2&  44.3&  39.3&  37.5&  47.6&  40.6&  40.5\\
\#3&  19.8&  21.2&  23.8&  38.1&  28.2&  21.7\\
field& 39.7&  40.5&  43.5&  64.2&  49.9&  42.6\\
\hline
Lup.3 & &&&&&\\
\#1& 16.6&  17.8&  19.4&  26.2&  21.5&  18.2\\
field& 15.7&  16.9&  18.6&  26.2&  20.9&  18.1\\
\hline
IC348 & &&&&&\\
\#1&  85.9& 108.9& 126.6& 154.7& 135.2& 111.9\\
\#2&   5.1&   2.3&   0.8&   2.4&   1.3&   3.1\\
field& 76.6&  74.8&  74.7&  83.3&  77.3&  78.2\\
\hline
\end{tabular}
\caption[]{Inference of $\cal M$ employing the distribution-by-number methodology in the stellar associations by \cite{KM11},  according different scenarios.}
\label{tab:table2}
\end{table}

\subsubsection{The probabilistic methodology}
\label{subsect:probmeth}

In the probabilistic case, pdfs are only used to describe unknown data, and observed data is used to define constraints over such unknown data, so that both types of data have different roles. In addition, the solution  cannot be summarized in a single value, but as a distribution function. Although some summaries of such distribution (as the mean value) can be obtained analytically, such values do not necessarily have enough  information, and the best method is to obtain the full distribution of possible solutions and work with it.
We propose here the methodology to obtain the probability distribution of $\cal M$ when we know the individual masses of the most massive $N_\mathrm{a}$ stars, and we know that all stars equal to or more massive than $m_\mathrm{a}^\mathrm{obs}$ are included in the $N_\mathrm{a}$ set. The problem cannot be solved analytically since recursive convolutions involving power laws (such as the IMF) have no analytical solution. So we can only propose the following step-by-step procedure:

\begin{enumerate}
\item {\it Obtain the distribution of $\cal N$,  $\Phi_{\cal N}({\cal N}|N_\mathrm{a})$, which can be inferred from the data using Eq.\ref{eq:phiNNa}}. We stress again that an assumption about $\Phi_{\cal N}({\cal N})$ is required. We note that the result would be quite dependent on the lower limit assumed in the $\Phi_{\cal N}({\cal N})$ distribution.

\item {\it Compute the distribution of $\Phi_{M_\mathrm{not-obs}}({M}_\mathrm{not-obs}| N_i)$ for the possible values of $N_i = {\cal N} - N_\mathrm{a}$ values obtained from the previous distribution}. The distribution provides the distribution of possible values of the total mass  {\it from the unknown stars}, ${M}_\mathrm{not-obs}$, that is, ${\cal M}$  is actually constrained to the non-observed stellar masses $m < m_\mathrm{a}^\mathrm{obs}$, so we must use a constrained IMF to describe what we do not know, $\phi(m| m < m_\mathrm{a})$. 

Such $\Phi_{M_\mathrm{not-obs}}({M}_\mathrm{not-obs}| N_i)$ distributions can be computed either by Monte Carlo simulations or by a numerical self-convolution.

\item {\it Compute the distribution of $\Phi_{\cal M}({\cal M}| M_\mathrm{a}, N_\mathrm{a})$}. This is done by weighting the previous  $\Phi_{M_\mathrm{not-obs}}({M_\mathrm{not-obs}}|   N_i)$ distributions by the probabilities of each $N_i$ value provided by  $\Phi_{\cal N}({\cal N}|N_\mathrm{a})$ and including the contribution to the total mass of the observed stars. 

\end{enumerate}

We note that these two last steps can be done by means of Monte Carlo simulations, which sample the {\it discrete} distribution $\Phi_{\cal N}({\cal N}|N_\mathrm{a})$ to obtain different $N_i$ values, and by sampling the constrained  IMF with this number of stars.
The previous procedure covers scenarios 2 and 3 by applying only step 2: obtain $\Phi_{\cal M}({\cal M}| {\cal N})$ or $\Phi_{\cal M}({\cal M}_\mathrm{not-obs}| N_i)$ for a known ${\cal N}$.

We applied this methodology to the set of clusters of \cite{KM11} under different scenarios by means of Monte Carlo simulations. The distribution of solutions for each cluster in each scenario was sampled by $10^7$ Monte Carlo simulations, and the resulting distribution was binned in intervals with $\Delta {\cal M} = 0.5 \mathrm{M}_\odot$. We note that in scenario 4 the simulations sample both the IMF and the assumed $\Phi_{\cal N}({\cal N})$ distributions (power laws with $\beta=0$ and $\beta=2$). Therefore, the simulations span a larger $\cal M$ range and an additional uncertainty is expected for the confidence interval estimations.

\begin{table}
\begin{tabular}{l|r|r|rr|rr|r|}
 \hline
 & \multicolumn{6}{c|}{$\cal M$ inferred in scenario 2} & \\
name&mean&mode&\multicolumn{2}{c|}{95.4\% CL}&\multicolumn{2}{c|}{68.3\% CL}&${\cal M}$ obs\\
 \hline
 Tau. & & &  & & & & \\
\#1 &    9.2&   5.9&   2.7&  21.2&   3.7&   9.7 &   10.6\\
\#2 &   13.8&   9.6&   4.8&  29.8&   6.8&  14.8 &   15.5\\
\#3 &    8.8&   5.2&   2.5&  20.5&   3.5&   9.0 &    8.1\\
\#4 &   11.1&   7.3&   3.5&  24.5&   5.0&  11.5 &   22.7\\
\#5 &    6.5&   3.7&   1.5&  15.5&   2.5&   7.0 &    8.2\\
\#6 &   14.3&  10.1&   4.9&  30.9&   6.9&  14.9 &   17.7\\
\#7 &   11.1&   7.3&   3.5&  24.5&   5.0&  11.5 &   16.1\\
\#8 &    7.4&   4.5&   1.7&  17.2&   2.7&   7.7 &   12.3\\
field &   80.3&  67.7&  47.5& 132.0&  55.5&  87.0 &   88.5\\
 \hline
 ChaI & & & & & & & \\
\#1 &    5.5&   2.9&   1.2&  13.7&   1.7&   5.7 &    3.7\\
\#2 &   44.3&  35.0&  22.8&  80.8&  27.8&  47.8 &   40.5\\
\#3 &   19.8&  13.6&   7.8&  40.8&  10.3&  20.8 &   21.7\\
field &   39.7&  31.0&  19.8&  73.8&  24.3&  42.8 &   42.7\\
 \hline
 Lup.3 & & & & & & & \\
\#1 &   16.6&  11.8&   6.0&  35.0&   8.5&  18.0 &   18.2\\
field &   15.7&  10.7&   6.0&  33.5&   8.0&  17.0 &   18.2\\
 \hline
 IC348 & & &  & & & & \\
\#1 &   85.8&  71.0&  51.2& 140.2&  60.2&  93.2 &  111.9\\
\#2 &    5.1&   2.8&   1.0&  12.5&   1.5&   5.0 &    3.1\\
field &   76.7&  64.3&  45.1& 127.6&  52.6&  83.1 &   78.2\\
 \hline
  \end{tabular}
\caption[]{Inference of $\cal M$ employing probabilistic methodology for the stellar associations by \cite{KM11} in scenario 2, using the observed value of $\cal N$.}
\label{tab:table3}
\end{table}

Table~\ref{tab:table3} shows the resulting mean, mode, and 68.3\% (equivalent to $1\sigma$ in a Gaussian distribution) and 95.4\%  (equivalent to $2\sigma$ in a Gaussian distribution)  confidence intervals around the mode for scenario 2. As expected, the mean value of the distribution coincides with the result of Eq.\ref{eq:Mdet} shown in Table~\ref{tab:table2}. All observed $\cal M$ are in the 94.5\% confidence interval around the mode, although only 27\% are in the 68.3\% confidence interval, being the observed $\cal M$ larger than the range quoted in such interval.

\begin{table}
\begin{tabular}{l|r|r|rr|rr|r|}
 \hline
 & \multicolumn{6}{c|}{$\cal M$ inferred in scenario 3} & \\
name&mean&mode&\multicolumn{2}{c|}{95.4\% CL}&\multicolumn{2}{c|}{68.3\% CL}&${\cal M}$ obs\\
 \hline
 Tau. &&&&&&&\\
\#1 &    9.7&   9.9&   8.6&  10.6&   9.1&  10.1&   10.6\\
\#2 &   14.7&  14.4&  13.7&  16.2&  14.2&  15.7&   15.5\\
\#3 &    7.8&   7.6&   6.8&   8.8&   7.3&   8.3&    8.1\\
\#4 &   22.7&  22.5&  21.8&  23.3&  22.3&  23.3&   22.7\\
\#5 &    8.7&   8.8&   8.0&   9.5&   8.0&   9.0&    8.2\\
\#6 &   17.7&  17.4&  16.7&  19.2&  16.7&  18.2&   17.7\\
\#7 &   15.9&  15.6&  14.9&  16.9&  15.4&  16.4&   16.1\\
\#8 &   12.2&  11.9&  11.2&  13.2&  11.7&  12.7&   12.3\\
field&   90.2&  90.2&  87.9&  92.9&  88.9&  91.4&   88.5\\
 \hline
 ChaI &&&&&&&\\
\#1 &    3.5&   3.4&   2.6&   4.6&   3.1&   4.1 &    3.7\\
\#2 &   39.3&  39.5&  37.3&  41.8&  38.3&  40.8&   40.5\\
\#3 &   21.2&  21.0&  19.7&  22.7&  20.2&  21.7&   21.7\\
field&   40.5&  40.6&  38.4&  42.4&  39.4&  41.4&   42.7\\
 \hline
 Lup.3 &&&&&&&\\
\#1 &   17.8&  17.5&  16.3&  19.3&  17.3&  18.8 &   18.2\\
field&   16.9&  16.9&  15.7&  18.2&  16.2&  17.7 &   18.2\\
 \hline
 IC348&&&&&&&\\
\#1 &109&109&106&112&108&111&112\\
\#2 &    2.3&   2.2&   1.4&   3.4&   1.9&   2.9 &    3.1\\
field&   74.8&  74.7&  71.9&  77.9&  73.4&  76.4 &   78.2\\
 \hline
 \end{tabular}
\caption[]{Inference of $\cal M$ employing probabilistic methodology for the stellar associations by \cite{KM11} in scenario 3, using the observed value of $\cal N$, $N_\mathrm{a}$, $M_\mathrm{a}$ and $m_\mathrm{a}=0.5 \mathrm{M}_\odot$.}
\label{tab:table4}
\end{table}

Table~\ref{tab:table4} shows the results of the $\cal M$ distribution for scenario 3, which includes a larger amount of information. The mean and mode of the distribution coincides (hence the distribution is symmetric), and the mean value is also coincident to the result of Eq.~\ref{eq:MdetfromNMyNtot}, as expected. However, in this case we can evaluate how good  this estimation actually is (and hence the distribution by number estimation). Taking favorable round-around cases, 17\% of the clusters (i.e., field stars in ChaI, IC348 \#1, and field stars in IC348) are outside the $2\sigma$ range, 83\% are in the $2\sigma$ range, and 67\% are in the $1\sigma$ range (i.e., 12 clusters). Given the low number of clusters for this study, we find this result partially consistent with a standard methodology. However, in theory, we would expect only one cluster outside the $2\sigma$ range, although we can invoke the use of a low number of clusters for this study. An additional outcome of this study is that, although Eq.~\ref{eq:MdetfromNMyNtot} produces results similar to the  observations, it does not necessarily provide a fully compatible (e.g., at $1\sigma$ level) result. Again, this enforces the idea of using the whole pdf of possible solutions instead a summary (like the confidence interval range) of it.

\begin{table}
\begin{tabular}{l|r|r|rr|rr|r|}
 \hline
  & \multicolumn{6}{c|}{$\cal M$ inferred in scenario 4 with $\Phi_{\cal N}({\cal N}) = \mathrm{cte}$} & \\
name & mean & mode & \multicolumn{2}{c|}{95.4\% CL} &  \multicolumn{2}{c|}{68.3\% CL} & ${\cal M}$ obs\\
 \hline
 Tau. & & &  & & & & \\
\#1 &   14.2&  13.4&   9.7&  19.7&  11.2&  16.2 &   10.6\\
\#2 &   21.1&  20.1&  15.4&  27.4&  17.4&  23.4 &   15.5\\
\#3 &   12.5&  11.8&   8.0&  18.0&   9.5&  14.5 &    8.1\\
\#4 &   35.6&  34.8&  28.5&  43.5&  31.5&  39.0 &   22.7\\
\#5 &   16.0&  15.0&  10.8&  21.8&  12.8&  18.3 &    8.2\\
\#6 &   25.7&  24.7&  19.4&  32.4&  21.9&  28.4 &   17.7\\
\#7 &   24.2&  23.1&  18.4&  30.9&  20.4&  26.9 &   16.1\\
\#8 &   14.6&  14.0&  10.7&  18.7&  12.2&  16.2 &   12.3\\
field&  127& 126& 112& 142& 119& 134 &   89\\
 \hline
 ChaI & & & & & & & \\
\#1 &    3.9&   3.0&   1.7&   7.2&   1.7&   4.7 &    3.7\\
\#2 &   41.3&  40.3&  34.1&  49.6&  36.6&  44.6 &   40.5\\
\#3 &   28.9&  28.4&  22.2&  36.2&  24.7&  31.7 &   21.7\\
field&   50.6&  49.5&  41.7&  59.7&  45.7&  54.7 &   42.7\\
 \hline
 Lup.3 & & & & & & & \\
\#1 &   22.2&  21.4&  16.7&  28.2&  18.7&  24.7 &   18.2\\
field&   21.7&  20.9&  16.1&  27.6&  18.1&  24.1 &   18.2\\
 \hline
 IC348 & & &  & & & & \\
\#1 &  136& 135& 122& 150& 128& 142 &  112\\
\#2 &    2.0&   0.8&   0.5&   4.5&   0.5&   2.5 &    3.1\\
field&   78.0&  76.9&  68.2&  88.7&  72.2&  82.7 &   78.2\\
 \hline
\end{tabular}
\caption[]{Inference of $\cal M$ employing probabilistic methodology for the stellar associations by \cite{KM11}, using the value of $N_\mathrm{a}$, $M_\mathrm{a}$ and $m_\mathrm{a}=0.5 \mathrm{M}_\odot$ and assuming a flat $\Phi_{\cal N}({\cal N})$ distribution.}
\label{tab:table5}
\end{table}

\begin{table}
\begin{tabular}{l|r|r|rr|rr|r|}
 \hline
  & \multicolumn{6}{c|}{$\cal M$ inferred in scenario 4 with $\Phi_{\cal N}({\cal N}) \propto {\cal N}^{-2}.$} & \\
name & mean & mode & \multicolumn{2}{c|}{95.4\% CL} &  \multicolumn{2}{c|}{68.3\% CL} & ${\cal M}$ obs\\
 \hline
 Tau. & & &  & & & & \\
\#1 &   12.7&  11.8&   8.6&  17.6&  10.1&  14.6 &   10.6\\
\#2 &   19.6&  18.8&  14.5&  25.5&  16.0&  21.5 &   15.5\\
\#3 &   11.1&  10.2&   6.9&  15.9&   8.4&  12.9 &    8.1\\
\#4 &   34.1&  33.4&  27.7&  41.7&  30.2&  37.2 &   22.7\\
\#5 &   14.6&  13.7&  10.0&  20.0&  11.5&  16.5 &    8.2\\
\#6 &   24.2&  23.5&  18.2&  30.7&  20.7&  26.7 &   17.7\\
\#7 &   22.7&  22.1&  17.4&  28.9&  19.4&  25.4 &   16.1\\
\#8 &   13.2&  12.5&  10.2&  16.7&  10.7&  14.2 &   12.3\\
field&  125& 124& 111& 140& 118& 132&   89\\
 \hline
 ChaI & & & & & & & \\
\#1 &    2.6&   2.0&   1.7&   4.7&   1.7&   3.2 &    3.7\\
\#2 &   39.9&  39.0&  32.7&  47.7&  35.7&  43.2 &   40.5\\
\#3 &   27.5&  26.6&  21.3&  34.3&  23.8&  30.3 &   21.7\\
field&   49.2&  48.2&  41.0&  58.5&  44.0&  53.0 &   42.7\\
 \hline
 Lup.3 & & & & & & & \\
\#1 &   20.7&  19.9&  15.7&  26.2&  17.7&  23.2 &   18.2\\
field&   20.2&  19.4&  15.2&  25.7&  17.2&  22.7 &   18.2\\
 \hline
 IC348 & & &  & & & & \\
\#1 &  135& 133& 121& 149& 127& 141&  112\\
\#2 &    0.8&   0.8&   0.5&   2.0&   0.5&   1.0 &    3.1\\
field&   76.6&  76.0&  66.7&  86.7&  71.2&  81.2 &   78.2\\
 \hline
\end{tabular}
\caption[]{Inference of $\cal M$ employing probabilistic methodology for the stellar associations by \cite{KM11}, using the value of $N_\mathrm{a}$, $M_\mathrm{a}$ and $m_\mathrm{a}=0.5 \mathrm{M}_\odot$ and assuming a power-law $\Phi_{\cal N}({\cal N})$ distribution with $\beta = 2$.}
\label{tab:table6}
\end{table}

Tables~\ref{tab:table5} and \ref{tab:table6} show the results of applying this methodology using flat and power law $\Phi_{\cal N}({\cal N})$ distributions ($\beta=0$ and $\beta=2$, respectively) in the range from ${\cal N}=N_\mathrm{a}$ to ${\cal N}=4000$ stars. The first result is that mean and mode values of the distribution are not equal in general, and the distribution is not symmetric, but j-shaped. The mode in the case of a flat $\Phi_{\cal N}({\cal N})$ distribution is similar to the result obtained by Eq.~\ref{eq:MdetformNyM}. In this case, the observed $\cal M$ of seven clusters are outside the $2\sigma$ confidence interval (actually, the clusters with lower $p_\mathrm{nor}(N_\mathrm{a}|{\cal N})$ value quoted before and ChaI\#3). If we neglect the six clusters with the larger deviations from the IMF, we obtain  a result showing that 9\% of the cluster are outside the $2\sigma$ interval, 91\% of the cluster are in the $2\sigma$ interval, and 55\% of the clusters are in the $1\sigma$ interval. This is a reasonable result of any statistical test.

Finally, the results of Eqs.~\ref{eq:MdetformM} and \ref{eq:MdetformNyM} are within the $2\sigma$ range in the case of a flat $\Phi_{\cal N}({\cal N})$ distribution, but the results of Eq.~\ref{eq:MdetformN}
(estimation from the extrapolations of the observed $N_\mathrm{a}$) produce larger values than the upper limit of  $2\sigma$.

\section{Discussion}
\label{sec:discusion}

We have shown in this work that the determination of cluster masses is not so trivial as  supposed in the literature. 
The distribution-by-number methodology uses known data to {\it determine} unknown data, whereas the probabilistic methodology uses known data to {\it constrain} unknown data. The problem is also related to the trade-off between unknown data and probability. When we use a pdf, like the IMF, to make inferences about unknown data, we implicitly renounce obtaining actual values of the inferred quantity. The price is to renounce precision in favor of accuracy. In contrast, the distribution-by-number methodology favors precision and renounces accuracy.
The difference is in the algebra (and the logic reasoning) used in each of the methodologies to manipulate formulae. The 
distribution-by-number methodology uses standard algebra, where symbols are just mathematical expressions without added meaning. The probabilistic methodology follows the algebra of probability, which implies a clear identification of the known and the (random) variables we aim to describe by a probability distribution.

As an example, the equation 

\begin{eqnarray}
{\cal N} \times  p(m \geq m\mathrm{_{a}^{obs}})  &=& N_\mathrm{a} \nonumber
\end{eqnarray}

\noindent provides an {\it estimation}Ê~of the number of stars with mass equal or larger than $m\mathrm{_{a}^{obs}}$ in a cluster with $\cal N$ stars. But such an estimation is not necessarily a mean value nor a mode value (cf. Paper I for the case that $N_\mathrm{a} = 1$). 
In that case, we know $\cal N$; hence, we are working with a $\Phi_{N_\mathrm{a}}(N_\mathrm{a}|{\cal N})$ distribution.
The inversion of the equation, that is,

\begin{eqnarray}
{\cal N} = \frac{  p(m \geq m\mathrm{_{a}^{obs}}) }{N_\mathrm{a}}, \nonumber
\end{eqnarray}

\noindent provides the modal value ${\cal N}^\mathrm{mode}$ of the distribution $\Phi_{\cal N}({\cal N}| N_\mathrm{a})$ when a flat distribution of $\cal N$  values is assumed, (i.e., $\Phi_{\cal N}({\cal N}) = \mathrm{constant})$. The distribution $\Phi_{\cal N}({\cal N})$ appears naturally when the Bayes' theorem is used. This is a natural result when we realize that, since $\cal N$ is  unknown, we need its probability distribution to  make inferences about it, and that the ``innocent" algebraic manipulation we have done has a completely different meaning than the one we would expect.

\subsection{To $\Phi_{\cal N}({\cal N})$ or not to $\Phi_{\cal N}({\cal N})$?}

We are now in the uncomfortable situation of having to assume a $\Phi_{\cal N}({\cal N})$ distribution in the inference of $\cal N$ and $\cal M$. However, the relevance of the $\Phi_{\cal N}({\cal N})$ in the inference of $\cal M$ is also dependent on the value of $m_\mathrm{a}$ and $N_\mathrm{a}$. In a back-of-the-envelope argument, the effect of a power-law $\Phi_{\cal N}({\cal N})$ distribution is to decrease $N_\mathrm{a}$ by $\beta$ stars (cf. Eq.~\ref{eq:Nmode} used for ${\cal N}^\mathrm{mode}$ estimation). Hence, the larger $N_\mathrm{a}$, the lower the dependence of the $\cal M$ estimation on $\Phi_{\cal N}({\cal N})$. Of course, the way to increase $N_\mathrm{a}$ is to be complete down to the lowest $m_\mathrm{a}$ possible.

In the cases where the $\cal M$ inference strongly depends on our choice of $\Phi_{\cal N}({\cal N})$, we must be guided by our knowledge of the physical system environment and the scientific goal of the analysis.
A flat $\Phi_{\cal N}({\cal N})$ assumes that there is no previous knowledge about the system {\it environment}, so it looks like good option
in the case of isolated systems and when we are only interested in the system properties.

However, the situation varies if 
we are interested in a cluster that  {\it we know}  is in a supercluster environment or is the result of 
molecular cloud fragmentation. In these cases, depending on our knowledge and hypothesis about star formation (SF), we can consider that such fragmentation is the result of a high-order structure; hence, the particular cluster is not an isolated entity. This would imply that some values of $\cal N$ or $\cal M$ are more probable than others, and this information must be taken into account in the inference of $\cal N$ and $\cal M$ of the particular cluster.

We must stress here that the proposed method only applies to $\Phi_{\cal N}({\cal N})$ distributions, and not to $\Phi_{\cal M}({\cal M})$ ones. The case of $\Phi_{\cal N}({\cal N})$ is easily implemented as far as it is related to sampling theory and the number of the elements in the sample is the relevant quantity. The inclusion of $\Phi_{\cal M}({\cal M})$ is not so trivial, since it depends implicitly on a $\Phi_{\cal N}({\cal N})$ distribution. However,  such distribution can not be obtained analytically (the convolution problem is not analytic in general cases). In addition, since $\Phi_{\cal N}({\cal N})$ is a discrete distribution, we have a large, but finite (and hence computable), number of cases. This is not true for $\Phi_{\cal M}({\cal M})$ because it is a continuous function and the possible solutions that a combination of $\cal N$ stars produces a particular $\cal M$ is infinite. At this moment, the only solution is to use $\Phi_{\cal M}({\cal M})$ as a proxy for $\Phi_{\cal N}({\cal N})$, which would be valid for situations where we know a priori that the minimum possible number of stars is large (i.e., $N_\mathrm{a}$ is large, or we have additional information about a minimum number of stars in the cluster).

Finally, the situation also changes if we are interested in obtaining $\Phi_{\cal N}({\cal N})$ or $\Phi_{\cal M}({\cal M})$ from a set of clusters. Following \cite{tarantola}, 
the most viable way is to make an iterative process. First, assume a $\Phi_{{\cal N},0}({\cal N})$ distribution and compute resulting distributions of ${\cal N}_i$ and ${\cal M}_i$ for each cluster. After that, combine such distributions to obtain {\it from the sample} the global distributions
$\Phi_{{\cal N},1}({\cal N})$ and $\Phi_{{\cal M},1}({\cal M})$. If $\Phi_{{\cal N},1}({\cal N}) \neq \Phi_{{\cal N},0}({\cal N})$, then $\Phi_{{\cal N},0}({\cal N})$ is not a self-consistent hypothesis. However, we must be aware that this does not prove that $\Phi_{{\cal N},1}({\cal N})$ and $\Phi_{{\cal M},1}({\cal M})$ are self-consistent hypotheses! The only way to achieve a self-consistent hypothesis is iterate the process until $\Phi_{{\cal N},j-1}({\cal N}) = \Phi_{{\cal N},j}({\cal N})$ being the $j-1$ distribution is the one used as input and the $j$ distribution is the resulting one, along with testing if the resulting $\Phi_{{\cal M},j}({\cal M})$ distributions also obey such a condition (a cross validation). 
However, we stress again that such a cross-validation process is a requirement that depends on the $N_\mathrm{a}$ value and that for large enough $N_\mathrm{a}$ values, the resulting $\Phi_{\cal M}({\cal M}| M_\mathrm{a}, N_\mathrm{a})$ solution for the $\cal M$ distribution of a cluster is almost $\Phi_{\cal N}({\cal N})$ independent.

\section{Conclusions}
\label{sec:conclusions}

Throughout this work, we have  explicitly developed  the use of the IMF to obtain different physical parameters of stellar systems from  limited information. We made extensive use of the IMF as a pdf, which allowed us to make  proper use of probability theory and, in particular, the properties of sampling distributions (where the total number of stars in the system is included) and conditional probabilities.

We  studied the methodology to obtain the distribution of possible $\cal N$ and $\cal M$ values from the knowledge of the set of the most massive stars in the system. The result is dependent  on
the values of $m_\mathrm{a}$ and $N_\mathrm{a}$, and
on the hypothesis about the overall distribution of the number of stars in clusters $\Phi_{\cal N}({\cal N})$, including the limits of such distribution (especially the lower one).

\begin{acknowledgements}
We  acknowledge Kevin Covey for discussions of the similarities and differences of $\Phi_{\cal N}({\cal N})$ and $\Phi_{\cal M}({\cal M})$ and their implications in the modeling of clusters and galaxies, which have been very useful for this paper and related works. We acknowledge the suggestions of the referee, Peter Anders, which greatly improved the clarity of the paper. 
This work has been supported by the MICINN (Spain) through the grants AYA2007-64712, AYA2010-15081, AYA2010-15196, AYA2011Ð22614, AYA2011-29754-C03-01, AYA2008-06423-C03-01/ESP,  AYA2010-17631,
programs UNAM-DGAPA-PAPIIT IA101812 and CONACYT 152160, Mexico, and co-funded under the Marie Curie Actions of the European Commission (FP7-COFUND)
\end{acknowledgements}

\clearpage

\end{document}